\documentstyle[preprint,aps]{revtex}

\def\bea{\begin{eqnarray}}
\def\eea{\end{eqnarray}}
\begin{document}
\draft
\preprint{}
\title{Bohmian Time Versus Probabilistic Time}
\author{M. Abolhasani\footnote{E-mail:ablhasan@netware2.ipm.ac.ir}$^{1,2}$, and M. Golshani$^{1,2}$}
\address{$^1$Institute for Studies in Theoretical Physics and Mathematics
(IPM), P.O.Box 19395-5531, Tehran, Iran.}
\address{$^2$Department of Physics, Sharif University of Technology,
P.O.Box 11365-9161, Tehran, Iran.}
\date{\today}
\maketitle
\begin{abstract}
One of the basic peoblems of quantum cosmology is the problem of time. Various
solutions have been proposed for this problem. One approach is to use the
Bohmian time. Another Approach is to use the probabilistic time which was recently
introduced by Castagnino. We consider both of these definitions as generalizations
of a semi-classical time and compare them for a mini-super space.
\end{abstract}
\section{Introduction}
In quantum cosmology, the universe is described by a wave function $\psi$. This
wave function can be obtained as a solution of the Wheeler-DeWit equation (WDW)
with appropriate boundary conditions. This equation is the relevant Schr\"odinger
equation ($H\psi=i\hbar \frac{\partial \psi}{\partial t}$) which is obtained
from the classical theory, using the Dirac quantization prescription. Since the
general covariance of the classical theory gives the classical Hamiltonian as
a subsidiary condition ($H=0$), the solutions of the WDW equation  are time-independent.
But how can a time-independent wave function describe a dynamical universe?
Various solutions have been proposed for this  so-called {\it time-problem}:
internal time versus external time, semi-classical time, Bohmian time,
probabilistic time,...etc. Here we compare three of these times: the Bohmian time[1],
the probabilistic time[2] and the semi-classical time[3]. First, we give two different
definitions of the semi-classical time. Then, we introduce the Bohmian
time and the probabilistic time as generalizations of the two definitions of the
semi-classical time. Consequently, we calculate the rate of the expansion of
the universe in terms of these two times for a mini-super space. Finally, we compare these two times.

\section{semi-classical time}
Consider a simple system having a Lagrangian[4] :

\begin{eqnarray}
L=\frac{1}{2}m(\dot Q^2-V(Q))
\nonumber
\end{eqnarray}
In this Lagrangian,  there is no coupling between the kinetic energy and the
potential energy terms that mimic gravity. The momentum $P_Q$, conjugate to $Q$,
is equal to $m\dot Q$. Now consider a WKB solution of this problem associated
with the energy E

\begin{eqnarray}
\psi_{_{WKB}}^{(E)}(Q)=\frac{N}{\surd {S'(Q)}}e^{\frac{i}{\hbar}S(Q)}
\nonumber
0\end{eqnarray}
where $S'(Q)=${\Large$\frac{dS}{dQ}$}, $S(Q)$ being the classical Hamilton-Jacobi function obtained
from
\begin{eqnarray}
\frac{1}{2m}S'+mV=E
\nonumber
\end{eqnarray}
To define a suitable time parameter, we take advantage of the classical equation
\begin{eqnarray}
P_Q=m\frac{dQ}{dt}=S'(Q)
\end{eqnarray}
Since in the Copenhagen interpretation of quantum mechanics (1) is merely
applicable in the semi-classical limit, this time parameter is only defineable
in this limit. For complex systems, however, it is not necessary for all
degrees of freedom to have semi-calssical behaviour. Thus, e.g., for our
present universe, gravitational degrees of freedom have semi-classical behaviour,
where as other fields have quantum behaviour.
Here, one can use classical degrees of freedom to define the aforementioned
time parameters. One can even obtain a time-dependent equation for the quantum fields
from the time-independent WDW equation[3]. It seems clear that for the early universe,
where all fields had quantum behaviour, this particular time parameter is not
well-defined in the Copenhagen interpretation. It is for this reason, that some
people have considered time as a classical concept which is born in the semi-classical limit.
\par
Now one can write
\bea
|\psi_{_{WKB}}^{(E)}(Q)|^2dQ=const \frac{dQ}{S'(Q)}\propto\frac{dQ}{\dot Q}=dt
\eea
This means that the the probability $|\psi_{_{WKB}}^{(E)}|^2$ is larger in the Q-interval
where the classical system spends more time. Thus, we can use (2) to define
time in the semi-classical limit too. These two different time parameters, defined for
this simple system, coincide in the classical limit.
\par
The question is whether we can extend these definitions to the region where quantum
effects are not negligible and systems are not simple.

\section{The Bohmian time}
When we extend the relation (1), defined in its semi-classical limit, to the
quantum realm, we get the Bohmian equation of motion[5]. In fact, one of the
fundamental priciples of the Bohmian mechanics is this relation which connects
the canonical momentum with the derivative of the phase of the wave function.
In this way, one can define a path for the particle, where $t$ is the parameter
of the path. One can use this time parameter to calculate the average tunneling
time through a potential barrier[6], where there is no well-defined way for its
calculation in the Copenhagen interpretation. In fact, one can design experiments
by which one can test the validity of the Bohmian time in the quantum domain[7].
\par
In recent years, people have used the Bohmian time to take care of the time
problem in quantum cosmology. To show this, consider the following mini-super space
\bea
ds^2=-N(t)^2dt^2+a(t)^2d\Omega_{3}^2
\nonumber
\eea
where $a(t)$ is the radius of the universe, $N(t)$ is an arbitrary function
of $t$  and $d\Omega_{3}^2$ is the metric of a unitary three-sphere. The
Hilbert-Einstein Lagrangian plus the Lagrangian of a homogeneous scalar field
$\phi$ is given in this metric by
\bea
L=-a^{-3}\ \left \{\ \frac{1}{2N} {\left [\ \frac{\dot a}{a} \right ]\ }^2+ \frac{N}{2} (-a^{-2}+H^2)
\right \} + a^3 \left \{\ \frac{\dot\phi^2}{2N} -NV(\phi) \right \}
\nonumber
\eea
where $V(\phi)$ is an arbitrary potential and $H^2$ is related to the cosmological
constant through the relation $H^2=\frac{\Lambda}{3}$. The canonical momenta $P_\phi$
and $P_a$ are, repectively, given by $a^3\dot\phi$ and $-a\dot a$. If we obtain the
classical Hamiltonian and make the substitutions $P_\phi\rightarrow -i\frac
{\partial}{\partial\phi}$ and $P_a\rightarrow -i\frac{\partial}{\partial\phi}$,
we obtain the WDW equation in the following form (in the gauge $N=1$)
\bea
\left \{\ \left [\ \frac{1}{2}a^{-3}(a\frac{\partial}{\partial a})^2+a^3(-a^{-2}+H^2)\right ]\ + \left [\ \frac{1}{2}
a^{-3}\frac{\partial^2}{\partial\phi^2}+a^3V(\phi) \right ]\ \right \} \psi(a,\phi)=0
\eea
Writing the solutions of this equation in the form $R(a,\phi)e^{\frac{i}
{\hbar}S(a,\phi)}$, we get the following Bohmian equations of motion
\bea
\dot a=-\frac{1}{a}\frac{\partial S}{\partial a}
\eea
\bea
\dot\phi=\frac{1}{a^3}\frac{\partial S}{\partial\phi}
\eea
Thus, knowing $\psi$, one can get the evolution of $a$ and $\phi$. The time
independence of $\psi$ for this problem (quantum cosmology) simplifies the
integration of (4) and (5).
\section{The probabilistic time}
Recently, Castagnino has extended (2) to define time in the quantum cosmology.
Consider the aforementioned mini-super space. The volume element is $\surd
{-G(a)} da d\phi$, where $G(a)=det(G_{ab})$, $G_{ab}$ being the metric defined
on the mini-super space. The probability of finding the metric in the interval
(a,a+da), independent of $\phi$, is
\bea
dP=da\surd {-G(a)} \int|\psi(a,\phi)|^2 d\phi
\nonumber
\eea
The idea of the probabilistic time is that the universe stays in metric $a$
for a period of time proportional to $dP$. Thus, we can define an element
of the probabilistic time in the following way
\bea
d\theta=c\ da\surd {-G(a)}\int|\psi(a,\phi)|^2 d\phi
\nonumber
\eea
where $c$ is a constant. Castagnino has considered a non-relativistic particle
described by the one-dimentional wave function $\psi(x,t)$. Suppose that we
parametrize $\psi$ by $\tau=\tau(t)$ instead of $t$, using an arbitrary measure
$\mu(\tau)$. Then, we can write
\bea
|\psi(x,t)|^2\  dx dt=|\psi_\mu(x,\tau)|^2 \mu(\tau)\ dxd\tau
\eea
Now, the question is about how we can obtain the real time $t$ from $\tau$. Using
(6), one can show that
\bea
dt=t_0\ d\tau\mu(\tau)\int|\psi(x,\tau)|^2 dx
\nonumber
\eea
where we have normalized $\psi(x,t)$ in the following way
\bea
\int_0^{t_0}\int|\psi(x,t)|^2 dx dt=1
\nonumber
\eea
For the problem under consideration, we can do the same thing with $a$ to obtain
\bea
d\theta=\theta_0\ da\surd {-G(a)} \int|\psi(a,\phi)|^2 d\phi
\eea
Using this relation, we can obtain the expansion rate of universe in terms of
the probabilistic time
\bea
\frac{da}{d\theta}={\left [\ \theta_0\surd {-G(a)}\int|\psi(a,\phi)|^2 d\phi\ \right ]\ }^{-1}
\eea
\section{Comparison of the Bohmian time and the Probabilistic time}
In the first section we showed that for a simple system, the Bohmian and the
probabilistic times coincide in the semi-classical limit. To compare these two
times in the quantum domain for the more complicated systems, we consider the
aforementioned mini-super space. In our discussion, we obtained the expansion
rate of the universe in terms of both the Bohmian time (4) and the probabilistic time
(8). The comparison of these two expansion rates provides a good way of
comparing these two time parameters. The relation (4) relates the expansion
rate to the phase of the wave function, where as the relation (8) relates the
expantion rate to the amplitude of the wave function. If we write $\psi(a,\phi)$
in the form $R(a,\phi)e^{\frac{i}{\hbar}S(a,\phi)}$ and substitute it in the WDW,
we get two equations, one of which is the following:
\bea
-a\frac{\partial}{\partial a}(R^2a\frac{\partial S}{\partial a})+
\frac{\partial}{\partial \phi}(R^2\frac{\partial S}{\partial \phi})=0
\eea
If $\psi$ is independent of $\phi$, this equation leads to:
\bea
R^2a\frac{\partial S}{\partial a} = const.
\nonumber
\eea
or
\bea
|\psi|^2=R^2=\frac{const}{a}(\frac{\partial S}{\partial a})^{-1}
\nonumber
\eea
If we insert this into (8), we get:
\bea
\frac{da}{d\theta}={\left [\ \theta_0\surd {-G(a)}\int\frac{const}{a}
(\frac{\partial S}{\partial a})^{-1} d\phi\right ]\ }^{-1}
\nonumber
\eea
Since $\psi$ was assummed to be independent of $\phi$, so is $S$. Thus, considering
the fact that for the mini-super space under consideration $\surd {-G(a)}=a^2$,
we get
\bea
\frac{da}{dt}=-\frac{1}{a}\frac{\partial S}{\partial a}
\nonumber
\eea
where we have defined $t$ as $\theta[-\theta_0 (const.)\int d\phi]^{-1}$.
\par
Notice that only in the non-realistic case of a universe free of matter,
the Bohmian time coincides with the probabilistic time. Here, we have
not referred to the semi-classical limit. In fact, for a system with one degree
of freedom, the Bohmian time and the probabilistic time coincide, both in
semi-classical regime and in the quantum domain as we have already shown. But,
for a complicated system, the form of (9) does not allow these two times to
coincide -- either in the semi-classical regime or in the quantum domain. Now,
the important question is about the relative merit of these two times. Equation (8)
indicates that the probabilistic time gives the expantion rate of the universe
independent of the amount of matter --something quite unnatural-- where as the
expansions rate in terms of the Bohmian time depends on the amount of matter in
the universe. On the other hand, while the Bohmian time reduces to the classical
time in the semi-classical limit no matter what the degrees of freedom of system
is, the probabilistic time coincides with the Bohmian one in the semi-classical
limit only when the degrees of freedom of the system is one. So, only in this
case, it reduces to the classical time. Therefore, the probabilistic time is not
a suitable parameter for the description of a dynamical universe.
\section{Conclusion}
If we don't accept the philosophy that time is a semi-classical concept,
then both the Bohmian time and the probabilistic time are more suitable definitions
for the time parameter than  the semi-classical time. Here, we have shown
that the probabilistic time has some problems that are not present for the
Bohmian time. For example, the rate of the expansion of the universe depends on
the amount of matter present in it, if it is expressed in terms of the Bohmian
time. But the same rate, is independent of amount of matter, if it is expressed
in terms of the probabilistic time.
\par
Furthermore, the Bohmian time reduces naturally to the semi-classcal time,
where as the probabilistic time has this property only for simple systems.

\end{document}